# Increase Investment in Accessible Physics Labs: A Call to Action for the Physics Education Community

July 29, 2021

Endorsed by the AAPT Board of Directors on January 5, 2022

**Dimitri R. Dounas-Frazer (he/him)**
Assistant Professor
Western Washington University

**L. C. Osadchuk**
M.A. Candidate
Western Washington University

**Daniel Gillen (he/him)**
Alum (B.S., 2017)
Haverford College

**Jamie Principato Crane (she/her)**
Independent Consultant

**Catherine M. Herne (she/her)**
Associate Professor
State University of New York at New Paltz

**Tyler M. Pugeda (he/him)**
Alum (B.S., 2019)
Johns Hopkins University

**Erin Howard (they/them)**
Undergraduate Student
Western Washington University

**Kevauna Reeves (she/her)**
Undergraduate Student
Washington State University

**Rebecca S. Lindell (she/her)**
Director and Owner
Tiliadal STEM Education

**Erin M. Scanlon (she/her)**
Assistant Professor in Residence
University of Connecticut

**G I. McGrew (he/they)**
Senior Instructor
Western Washington University

**David Spiecker (he/him)**
Ph.D. Student
Institute of Optics, University of Rochester

**J. Reid Mumford (he/him)**
Instructional Resource Advisor
Johns Hopkins University

**Sheila Z. Xu (she/her)**
Alum (B.S., 2014)
Massachusetts Institute of Technology

**Newton H. Nguyen**
Ph.D. Candidate
California Institute of Technology

## Acknowledgements

This report was written in response to a charge from the American Association of Physics Teachers (AAPT) Committee on Laboratories.

AAPT provided financial support through a Special Projects Grant, which funded honorariums for some of the authors.

The Western Washington University (WWU) Disability Access Center (DAC) provided resources for American Sign Language (ASL) interpretation during many meetings.

The National Technical Institute for the Deaf (NTID) provided resources for ASL interpretation during conference panels about this report at the 2021 AAPT Virtual Winter Meeting.

We gratefully acknowledge input from Dan P. Oleynik, Caroline Bustamante, Jacquelyn Chini, S. K. Wonnell, and Susanne Morgan.

## Recommended Citation: IEEE Format for Technical/Company Reports

# Table of Contents

# EXECUTIVE SUMMARY

Many physics laboratories prohibit learners from learning and engaging with physics in an equitable way. We call on the physics community to invest time, energy, and resources to increase accessibility of undergraduate physics labs.

The American Association of Physics Teachers (AAPT) Committee on Laboratories assembled a task force whose charge was to write an open letter to the physics education community calling for increased investment in accessible lab courses. Contributors to this paper include students, staff, and faculty with and without disabilities who expressed interest in the open letter. In this document, we recognize the need for making physics laboratories more accessible in all spaces (e.g., high school courses, graduate level courses, research labs). We focus on the experiences of students with disabilities in physics lab courses at the undergraduate level because that is the context for which the writing team had the most collective experience. The intended audiences for this document consist of undergraduate physics students, staff, and faculty, especially those who have direct stake in laboratory courses; physics departments; and member societies, including AAPT.

We begin by presenting our motivation for the document and the importance of accessibility and diversity in education and the workforce. We start with the broader context of accessibility, narrowing our focus to physics education and the current state of affairs and availability of accessible resources. Accessibility is then discussed in the specific context of physics laboratory courses, focusing on how barriers are created and can be lowered. In exploring ideas and strategies for improving accessibility, we recognize that the development of multiple pathways for laboratory investigation creates opportunities to expand learning opportunities for more students in physics lab programs.

We identify two types of investment, proactive and reactive investment. Strategies for increasing proactive investment are outlined for six target groups: physics lab instructors and staff, physics education researchers, physics departments, online physics content creators, physics conference planners, and member societies. While reactive investments play a crucial role in accessibility (i.e., providing accommodations for individual students' specific needs), the focus of this paper is on proactive investment strategies that benefit all learners. Neither type of investment eliminates the need for the other, but through proactive investment we can create lab courses that support a broader range of students and communicate the expectation of a diverse student body.

We hope that readers will reflect, share, collaborate, and take action toward making physics labs more accessible to all. We also hope that readers will share and discuss this document with others. Doing so normalizes conversations about disability, accessibility, and ableism/disablism in physics and other sciences.

# LIST OF IDEAS FOR INVESTMENT

**A. Physics lab instructors and staff**
- A-1. Incorporate principles from Universal Design for Learning (UDL) into the design and implementation of laboratory courses and activities.
- A-2. Include students, staff, and faculty with disabilities as partners in the design and implementation of accessible lab courses and activities.
- A-3. Communicate early and often about accommodations.
- A-4. Seek out and participate in professional development related to disability and accessibility.
- A-5. Align your actions with your beliefs.

**B. Physics education researchers**
- B-1. Incorporate theories of learning from Disability Studies into physics education research.
- B-2. Prioritize the experiences of students with disabilities within lab courses.
- B-3. Prioritize research and development of accessible lab courses.
- B-4. Move beyond the traditional researcher-participant dichotomy.
- B-5. Consider accessibility when communicating research findings.

**C. Physics departments**
- C-1. Regularly discuss which components of lab courses are fundamental.
- C-2. Describe each laboratory course comprehensively in course catalogs.
- C-3. Eliminate ableist/disablist language when teaching lab courses and when talking about them.
- C-4. Advise students appropriately.
- C-5. Pay for professional development of students, staff, and faculty.

**D. Online physics content creators**
- D-1. Establish and enforce accessibility guidelines for websites and content.
- D-2. Create and host blogs, forums, or listservs about accessibility in physics labs.
- D-3. Host or link to resources about accessibility in physics labs.

**E. Physics conference planners**
- E-1. Establish and enforce accessibility guidelines at conferences.
- E-2. Increase awareness of conference-related accessibility issues among organizers.
- E-3. Clearly & publicly communicate conference-related accessibility resources & constraints.
- E-4. Provide grants to defray increased travel costs for people with disabilities.
- E-5. Solicit post-conference feedback from attendees regarding accessibility.

**F. Member societies**
- F-1. Create and advertise professional development opportunities for students, staff, and faculty.
- F-2. Increase awareness of ableism/disablism and impacts on physics lab courses.
- F-3. Identify and advertise funding opportunities for improving accessibility of physics labs.
- F-4. Purchase assistive devices in bulk for reduced per-unit price and sell them at cost to physics departments and individual educators.

## 1. Charge, process, scope, and audience

According to its mission statement [1], the American Association of Physics Teachers (AAPT) Committee on Laboratories "values inclusive and accessible laboratory learning environments because learners, educators, and other stakeholders deserve the opportunity to learn, teach, and practice physics free from bullying, harassment, explicit or implicit bias, or **unmet accommodations**" (emphasis added). Further, the Committee will "write white papers, position statements, or policy documents in response to articulated community needs and interests." Consistent with these aspects of its mission, in January 2019, the Committee assembled a task force whose charge was to write an open letter to the physics education community calling for increased investment in accessible physics lab courses. In October 2019, the task force received funding through an AAPT Special Projects Grant to provide honorariums to students, staff, and faculty with disabilities who expressed interest in contributing to the ideation or writing of the open letter. Collectively, these stakeholders and the members of the task force formed a writing team who worked together to outline, draft, critique, and revise the document you are currently reading.

The total writing process covered more than a year of in-person and remote collaboration, initially consisting of topical working groups and in-person meetings supported by American Sign Language (ASL) interpretation. After a six-month writing hiatus due to the COVID-19 pandemic, project writing was resumed entirely remotely, with regular drop-in writing times for collaborative writing, editing, and discussion. Throughout this process, several additional collaborators joined the team, creating an opportunity to expand the scope of our original charge. Rather than write a brief open letter, we created a more comprehensive document that not only calls for increased investment in accessibility in physics labs, but also provides additional ideas and recommendations about what such investment could entail.

Meanwhile, in January 2020, the AAPT Committee on Laboratories and the Committee on Diversity agreed to co-sponsor two "Making Physics Labs More Accessible" panels at the 2021 AAPT Virtual Winter Meeting. One panel focused on perspectives of current physics students [2] and the other on those of former physical science students [3]. Several members of the writing team participated as moderators or panelists, making the panels an important introduction to, and extension of, this document. The purposes of the panels were to raise awareness about the white paper and provide the AAPT community with opportunities to communicate with members of the writing team about their motivations, goals, experiences, and recommendations.

Discussions about accessible labs within the Committee predate the formation of the task force. For example, at the 2018 AAPT Summer Meeting in Washington, D.C., the Committee on Laboratories and the Committee on Diversity in Physics co-sponsored a session called "Improving accessibility for physics students with visual or cognitive disabilities." Among the presentations in that session was a talk called "A Hands-on, Nonvisual Approach to Accessible Intermediate Physics Laboratory Courses," which was co-presented by Paul Thorman and Daniel Gillen [4]. Accordingly, this white paper is not the start of Committee efforts to improve accessibility, nor should it be the end of such efforts. Rather, this document calls on the physics education community to enhance investment in accessible labs.

People sometimes draw delineations between accessibility and inclusivity, where accessibility focuses solely on providing access to the learning environment (e.g., course content, physical spaces, interactions with peers), whereas inclusivity goes beyond access to ensure individuals are supported or planned for in the learning environment. In this report, we will not use this distinction because, in the past, it has been used as a mechanism to shift responsibility (e.g., from instructors to administrative staff) rather than to frame a more nuanced discussion of how change can be directly implemented by instructors. For example, in trying to make data acquisition software accessible, an instructor may believe that the task of checking the screen reader compatibility of the software would be the responsibility of staff in their local office of disability services. Defining screen reader compatibility as an 'accessibility practice' that is distinct from 'inclusive teaching' could result in an instructor neglecting to consider the screen reader compatibility of the software they choose unless required to do so by administrators. Thus, throughout the report we use the term 'accessibility' to encompass both access and inclusion, as traditionally defined, when describing practices and policies to improve the experiences of people with disabilities in physics labs.

We recognize the need for making physics laboratories more accessible for students, staff, and faculty with disabilities in high schools, colleges, universities, research facilities, and elsewhere. However, we chose to focus on the experiences of students with disabilities in lab courses at the undergraduate level, because that is the context for which the writing team possessed the most collective experience. The intended audiences for this document include undergraduate physics students, staff, and faculty, especially those who have direct stake in laboratory courses; physics departments; and member societies, including AAPT.

The document is organized as follows. First, we motivate the need for accessibility in education generally, physics education more specifically, and physics laboratory courses in particular. Connections between accessibility, creativity, and innovation are also discussed. Next, we highlight 27 ideas for investing in accessible physics labs for various stakeholders: lab instructors and staff, physics education researchers, physics departments, conference organizers, online content creators, and member societies. Finally, we conclude by encouraging readers to take action and make physics labs more accessible! In the appendices, we include anecdotes about the lived experiences of authors from their perspectives as current or former students (Appendix A), and we provide information about learners who are Autistic (Appendix B), blind or low vision (Appendix C), and/or Deaf or hard of hearing (Appendix D); some of the authors belong to one or more of these groups. In Appendix E, we provide a glossary of select terms related to accessibility and disability.

## 2. Motivation

### The need for accessible education

Before we can address accessibility in physics labs for people with disabilities, it is necessary to establish why accessibility is important. Reasons can range from legal compliance to strategies for promoting healthy, creative, and diverse collaborations within any industry. The cornerstone, however, is to realize that without giving adequate consideration to accessibility and providing reasonable accommodations when needed, we are overtly excluding people with disabilities.

First and foremost, providing reasonable accommodations for students and faculty with disabilities is required of educational institutions by law. In particular, there are three major federal laws that address the inclusion of people with disabilities in places of learning. These are the Rehabilitation Act of 1973, Individuals with Disabilities Education Act (IDEA) of 1975, and Americans with Disabilities Act (ADA) of 1990 [5]. Legislation is continuously evolving to keep up with changes in learning environments, technology, and population. Individual and class action lawsuits brought against institutions in recent decades continue to shape the legislation and accepted practices surrounding both higher education [6] and employment across all industries. Such lawsuits also inform efforts to improve representation of people with disabilities within the federal government in organizations that shape research, industry, and the policy governing both [7]–[9].

A significant part of the population has disabilities. Approximately 10% of 18-64 years olds in the United States have some sort of disability [10]. Just within AAPT and APS, about 1 in 5 members have one or more disabilities [11], [12]. It stands to reason that, to achieve a truly diverse community in any field, inclusion of people with disabilities is essential, and therefore, so is accessibility.

In order to increase diversity in any field or industry, it is necessary to prioritize accessibility in learning environments so that people with disabilities have equal access to resources for success and are fully included in the learning process. In physics and other science, technology, engineering, and mathematics (STEM) fields, higher education is almost always the entry point into the field. If students with disabilities do not feel a sense of belonging in undergraduate STEM programs, then they will be less likely to pursue a degree or career in a STEM field [13]–[15]. Thus, a focus on accessibility in undergraduate physics education is necessary (but not sufficient on its own) to improve representation of physicists with disabilities and make physics more accessible.

### The need for accessible physics education

Members of the physics and physics education communities must examine the type and quality of resources available to educators regarding accessibility in physics labs. As it stands, there exists a body of literature regarding accessibility specifically in physics [16]–[18]. However, those works are not well advertised, and they may be redundant or misaligned with each other. Recent work from Project ACCESSS at the University of Central Florida summarizes some of the resources available for accessible physics instruction and the types of disabilities those resources address [17]. Research has shown that faculty lack knowledge of accessibility laws [19]–[21], lack knowledge of how to support students with disabilities in their courses [22]–[24], and do not feel prepared to support students with disabilities in their classes [25], [26]. Furthermore, some of the findings demonstrate that active engagement strategies are not necessarily aligned with principles of Universal Design for Learning (UDL) [18]. For more information about UDL, we refer the reader to the glossary in Appendix E.

Adopting a UDL framework will make the classroom more accessible for students with disabilities and enrich the learning experience for all students. To enact practices that are aligned with UDL principles, physics educators would benefit from having examples or other forms of support [17]. We emphasize that adopting UDL practices will not eliminate the need to accommodate individual students with disabilities, but it should decrease the amount of additional accommodation that a teacher will need to provide to students. In [Section 3](), we elaborate on this idea by distinguishing proactive efforts (e.g., incorporating principles of UDL into the classroom) from reactive efforts (e.g., implementing specific accommodations for an individual learner).

Regarding the availability of resources discussing accessibility in physics, other disciplines—such as chemistry [27] and astronomy [28]—have detailed resources specific to the discipline. Although there is no comparable work for the discipline of physics in the United States, the Institute of Physics in the United Kingdom and Ireland has a guide to recommended practices for university physics departments [16], and a Canadian organization called Accessible Campus has resources for accessible science laboratories [29]. Additional resources can be drawn from organizations that focus on accessibility in science labs in general to guide us on making recommendations for physics labs in particular [30].

In order for resources about accessibility in physics labs to be useful, they need to be accompanied by professional support. Educators and administrators must be aware that the resources exist, and they must be able to adapt recommendations to their local contexts. Presently, the Advanced Laboratory Physics Association (ALPhA) immersions and Conferences on Laboratories Beyond the First Year of Instruction ("BFY Conferences") do not focus on disability, accessibility, universal design, or accommodations. Moreover, the AAPT New Faculty Workshop also does not address those topics. Apart from a 2001 Task Force on Physicists with Disability [31], no units in the American Physical Society (APS) are specifically dedicated to disability or accessibility. Clearly, there is a need for increased professional focus on, and support for, physicists with disabilities and corresponding issues related to access and inclusion.

The need for accessible physics laboratory courses

Improving physics labs is a national priority [32], [33] and some physics education researchers have specifically called on the physics community to invest in accessible and inclusive lab learning environments [34]:

"Labs use specialized equipment and software, and they often involve frequent peer-peer or student-instructor interactions. Therefore, labs may give rise to a unique combination of stereotypes, discriminatory behaviors, and mobility or sensory barriers that unfairly prevent full participation for some learners. Improved accessibility and inclusivity can be supported by research and development of labs that minimize barriers to students and educators from marginalized groups (e.g., people of color, people with disabilities, people who are lesbian, gay, bisexual, transgender, intersex, queer, or Two Spirit [LGBTIQ2S], women, and people from the intersections of these groups)."

Several accessibility issues are specific to labs. Our committee has been considering ways that the apparatus, learning goals, and social interactions in a laboratory setting can create unnecessary barriers for students with disabilities. In this process, we have relied heavily on the social model of disability, which frames disability as caused by the way society is organized. We also acknowledge the existence of the medical model of disability, which frames disability as caused by a person's impairments. For more information about these two models, please refer to the glossary in Appendix E.

In the following list, we highlight how disability and accessibility come into play at various stages of designing and teaching a physics lab course, from learning objectives through design of specific lab manuals:

- Learning objectives themselves might encode ableist/disablist ideas when they focus on the equipment being used rather than the essential functions of the equipment, such as the measurements being made. For example, stating that students should "learn to use an oscilloscope" might preclude access for individuals with visual and/or mobility impairments, and could be adapted to "learn to measure the amplitude, phase, and frequency of an AC signal; likely with an oscilloscope."

- Often, labs meet for longer periods of time per credit. The larger number of consecutive hours of class time can create barriers for people who experience fatigue.

- During labs, students often engage in intensive interaction with peers and instructors. Such interactions may unduly tax students who experience social anxiety or difficulty communicating in a social setting. Further, as the number and length of interactions increase, so, too, does the likelihood of (sometimes unintentional) interpersonal ableism/disablism, which can contribute to making students with disabilities feel excluded, unwelcome, or not useful.

- Equipment is typically not designed to be operated by people with limited mobility or vision. It may be large, heavy, and difficult to operate for people who have limited strength or dexterity. Screens and parts may require fine vision.

- Software and hardware may be inaccessible due to incompatibility with assistive technologies, such as screen readers or alternative keyboards.

- Safety concerns include that students who communicate visually or in other non-auditory ways may not hear audible warnings, and that service animals may be injured during an experiment.

- Instructions and activities for labs are often framed in the context which is typical for most students.They rarely or never acknowledge potential accommodations, which may not currently exist. For example, data and signals are typically visualized rather than sonified during collection, analysis, and reporting of data.

- Although instructors often intend to write open-ended labs for pedagogical purposes, that may be a disadvantage for some students who need specific step-by-step instruction to reach the intended learning outcome for the experiment.

- Lab manuals may also be inaccessible when they are not provided in electronic text-based formats and contain no alternative text (alt text) of images, which would be inaccessible to students using screen-reading software.

Laboratory courses are exciting teaching and learning environments, because teachers and students have opportunities to engage with advanced technology, apparatus, and software. Likewise, improving the accessibility of labs is also exciting because teachers and students can think comprehensively about multiple modes of engaging with that technology. In addition, making labs more accessible can include creatively rethinking how we incorporate existing, low-cost technologies into the classroom. Below, we list some examples of questions that teachers and students might ask themselves and each other when improving the accessibility of a lab course or activity:

- Which activities use technology for which there already exists commercially available accessible alternatives tailored to educational science contexts? Examples include talking LabQuest devices.

- How can the hardware or software in the lab be augmented with emergent assistive technologies? Examples include eye trackers or methods to detect muscle tensioning or electromyographic signals.

- How can low-cost home health care and monitoring devices be used as tools in the lab? Examples include vibrating toothbrushes, motorized blood-pressure cuffs, talking bathroom scales, and other personal devices that use voice as input or output.

- How can gaming devices be used as tools in the lab? Examples include devices that use gesture sensing, voice command, touch pads, virtual or augmented reality, or remote control.

- What could be achieved with a smart laboratory room whose array of sensors and actuators can couple to a student's digital notebooks? We imagine exciting applications of wireless sensor networks, home monitoring and control devices, QR codes, and physical objects embedded with sensors along lines of "internet of things" technology.

- How can we create versatile platforms for inclusive and multi-pathway learning in the lab using microcontroller systems (e.g., Arduino, Raspberry Pi, or Beaglebone) or programmable-system-on-chips?

Connections between accessibility, creativity, and innovation

The development of multiple pathways for laboratory investigation creates opportunities to expand the learning of all participants in a physics lab program. When an access need arises for one individual, other students—lab partners and even entire laboratory sections—can be enlisted into problem solving to meet those needs, if the individual learner consents to the process. (Note: It is not appropriate for an instructor, staff member, student, etc. to disclose students' disability and/or accommodations without the express consent of the student.) All students can collaborate in the design of adaptations to experimental apparatus and protocols. Such a process engages physical reasoning, technical skills, and interpersonal connection, all of which are relevant to future career prospects for physics students.

The impact of an adaptation of an experiment goes well beyond the individual whose access needs are addressed: new modes of perception and assimilation of experimental events are made available to all participants, deepening understanding and posing new questions. For example, Wanda Díaz Merced, a blind astronomer and computer scientist, has made significant advances in sonification of astronomical data, which makes the field of astronomy more accessible while providing a new way for all astronomers to engage with data [35]. Similarly, Steven Sahyun's work on auditory [36] and tactile [37] representations of graphs has important instructional implications for engaging all learners more deeply.

The current and ongoing COVID-19 pandemic has caused massive suffering and loss of life across the globe. It has significantly disrupted our everyday 'normal', oftentimes requiring novel adaptations to our lifestyles in both our living and working environments. Those of us who have been able to continue teaching, learning, and doing physics during the pandemic have made significant changes to our approaches to education and research in order to reduce transmission of the SARS-CoV-2 coronavirus. Public health mandates—such as social distancing and the sudden switch to remote teaching and learning—created an impetus to combine basic safeguards like wearing face masks with advanced technologies like internet-based virtual meetings. In response, many physics educators and departments have demonstrated the ability to change learning environments to simultaneously accommodate several modalities of engagement during a crisis.

Compared to traditional in-person modes of instruction, some learners found aspects of remote and hybrid environments to be more inclusive and accessible. Going forward, as we look ahead to the eventual receding of the pandemic, we have the opportunity and responsibility to continue rethinking and redesigning environments to improve their inclusivity and accessibility while resisting the temptation to return to an everyday 'normal' that has historically been exclusionary to some learners.

The engagement of students with many abilities to meet diverse needs goes well beyond the physics lab. Physics can be seen as a powerful tool for meeting human needs outside of the learning environment. Ideas that first are explored in a learning laboratory can translate into powerful solutions for broad needs around the world. We are reminded of the words of the late Zohra Aziza Baccouche (1976–2021), a blind nuclear physicist and filmmaker who aimed to connect scientific breakthroughs with the human experience. She once said, "We know power is work over time, that strength is endurance over time. So I endured a lot of obstacles, but at the same time I created strength and vision and wisdom and endurance" [38].

### 3. Ideas for investing in accessible physics labs

All students have different needs, skills, and interests that fall on a distribution, and students with disabilities are no exception. To accommodate as many students as possible, it is therefore necessary to make physics lab courses more accessible through both proactive and reactive investments.

**Proactive investment** is defined as unprompted changes made to increase accessibility. Proactive investment can include adopting a Universal Design for Learning (UDL) framework [39] or Universal Design of Instruction (UDI) [40] to make labs, classes, and activities accessible to as many students as possible, independent of the class demographics, and before the start of the lab, class, or activity. As an example, the University of Colorado Department of Astrophysics & Planetary Science has used UDL to improve introductory astronomy labs [41]. Proactive investment may entail the following:

- Requesting input and needs from the student at regular intervals.

- Forming an accessibility review committee that includes students, teaching staff, and disability services administrators that have funding and the ability to implement policy changes at the departmental or institutional level.

- Making mentors readily available to students with disabilities, because students can find it difficult to communicate their needs directly to faculty or staff. Mentors can be other students with disabilities or people who identify as not having a disability.

- Investigate who is supported or planned for within the current labs and who is not supported, planned for, or tasked with a focus on how the labs could be revised to provide support for all learners.

- In relation to the expectations for how students should engage with the lab (e.g., record data, physically interact with apparatus, coordinate analysis with their peers), interrogate how these expectations could preclude engagement for people with disabilities including those with physical/mobility, cognitive, health, visual, hearing, and mental health impairments. (Scanlon and Chini [42] developed a tool to examine physics courses from a disability perspective.)

**Reactive investment**, by contrast, is defined as changes made to increase accessibility after being prompted by an individual or a group seeking accommodations. As an example, students, staff, and faculty in the Department of Physics and Astronomy at Haverford College partnered with an accessibility consultant to accommodate a blind physics student [43]. Accommodations should be provided as quickly as possible upon request of the individual or group, and the people requesting accommodations should be regularly consulted on the effectiveness of changes made to apparatus or instruction. Reactive investment may entail the following:

- 3-D printing can enable a visually impaired user to obtain tactile information about the spatial distribution of atoms in a molecule.
- Purchasing an electric stirrer can enable a student with limited hand and arm dexterity to more independently complete a chemistry lab.
- Providing an American Sign Language (ASL) interpreter can make activities accessible to Deaf and hard-of-hearing individuals who use ASL.
- Allowing students extra time for completing lab write-ups.

Neither type of investment eliminates the need for the other. However, because reactive investments are designed to accommodate the specific needs of an individual student, the focus of this paper is on proactive investment strategies that benefit all learners. There are multiple stakeholders with access to material resources and decision-making authority relevant to proactively improving the accessibility of undergraduate physics labs, including funding agencies, professional societies, universities, offices of disability and accessibility, lawyers, interpreters, educators, education researchers, students, and so on. In this section, our intended audience is the following stakeholders:

A. Physics lab instructors and staff
B. Physics education researchers
C. Physics departments
D. Online physics content creators
E. Physics conference planners
F. Physics member societies

In the following subsections, we identify multiple ideas for investment tailored to each stakeholder listed above. We emphasize that each of these ideas should be discussed and implemented in partnership with physics students, staff, and faculty with disabilities.

A. Physics lab instructors and staff

Physics lab instructors and staff are in the unique position of having an opportunity to re-conceive physics instruction with multiple supports and options for learning incorporated at the outset. Likewise, they can examine the repertoire of laboratory technical competencies expected of students—instrumental, procedural, and design—and find multiple ways to develop, express, and assess these competencies. For example, the oscilloscope is considered an essential and versatile instrumental competency. The use of an oscilloscope in the lab can be expanded to be more accessible by investigating the following questions: How can capabilities for examining periodic waveforms, aperiodic but continuous waveforms, and pulses be captured and perceived by non-visual means? How can the options for adjusting signal gain, time resolution, triggering, or juxtaposition of multiple signals be accomplished without manual operation? Can the resulting solutions be incorporated as broadly useful enhancements to future commercial versions of these instruments? Answers to such questions enhance accessibility of the lab course.

When imagining the exciting landscape of possibilities for improving accessibility of experiments, physics lab instructors and staff should keep in mind that not all students know how to or feel empowered to advocate for themselves effectively in all contexts, so it is important for instructors to initiate discussions with students about accessibility and agree on mutually beneficial solutions. We acknowledge that not all instructors are necessarily knowledgeable about accessibility, which is a major reason for this paper. We recommend the following ideas for increasing accessibility for lab instructors and staff to consider:

A-1. **Incorporate principles from Universal Design for Learning into the design and implementation of laboratory courses and activities.** Instructors and staff should familiarize themselves with the Universal Design for Learning (UDL) framework and implement UDL-aligned instructional strategies to increase accessibility. Because there are many different strategies that instructors and staff may pursue, we share four select examples:

○ **Clearly describe all laboratory activities in advance.** For each lab course, ensure that enrolled students are aware of the technical, spatial, social, and sensory requirements of upcoming lab activities. Provide students and other stakeholders with copies of lab manuals and other course materials as early as possible. During the development of lab manuals and course materials, identify the fundamental human operations that are frequently used in conducting experiments along traditional lines, and to develop several alternatives to achieve an equivalent outcome. Examples of operations include manual assembly, adjustment of controls (e.g., switches, knobs), and visual reading of gauges and numeric displays. The fundamental human operations and alternatives should be included in the lab manual.

○ **Communicate with students redundantly across multiple platforms.** Important information about course content and logistics should be communicated to students in multiple places and in multiple formats (e.g., written in the lab manual, verbally during the lab period, diagram sent via the learning management system), it should be well organized, and it should be easy to reference throughout the duration of the course.

○ **Use and advertise multiple types of instructional materials.** Ensure that instructional materials are offered in a variety of formats (e.g., provide text handouts to accompany verbal lectures, provide digital versions of lab materials). Prior to the start of class, students should know which materials will be used, and they should have access to them. Also, consider leveraging different technologies and technological platforms to increase accessibility.

○ **Build in pauses or "wait time" between successive questions and activities.** Allow for different mental processing times by students. Some students need more or less time than others to process and internalize information. In addition, some students need to set aside time for ASL interpretation of verbal instructions, conversion of text to Braille or speech, coordinating with a note taker, repeating instructions, or some other communicative process that needs to happen first, before a student can grapple with the task at hand.

A-2. **Include students, staff, and faculty with disabilities as partners in the design and implementation of accessible lab courses and activities.** Create an independent study opportunity where students can receive credit or earn pay to collaboratively develop accessible physics lab activities in partnership with people who have disabilities and/or who are experts on accessibility. A common saying in disability justice movements is, "Nothing about us without us is for us;" see, for example, *Nothing About Us Without Us* by James Charlton (University of California Press; 2000). Put another way, any effort that is ostensibly about improving the experiences of people with disabilities must include people with disabilities in decision-making roles at all stages. Indeed, efforts to accommodate science students with disabilities without their input are harmful. Further, many physicists are uninformed about issues of disability and accessibility [45, 101] and therefore should not lead accessibility efforts on their own.

A-3. **Communicate early and often about accommodations.** When students identify particular accommodations, frequent communication between instructors, technical support staff, accessibility experts, and students is a key factor in implementing those accommodations in sensible ways, thus improving the accessibility of lab courses. If the expectations of lab courses and activities are clearly communicated prior to registration, students may be able to anticipate some accessibility needs, and conversations about accommodations can be initiated before the first day of class. Likewise, when planning an event, such as an activity or a meeting, consider the timing as to provide students with an opportunity to seek out additional resources or accommodations.

   ○ **Recognize that accommodations may be imperfect.** For example, some American Sign Language interpreters may not be fluent in signs for technical jargon, or they may not understand physics material well, which creates the possibility for miscommunication during ASL interpretation of spoken English. If instructors assume that interpretation is a perfectly efficient process, they might erroneously attribute a lack of understanding to the student rather than to the interpreter. Importantly, do not assume that miscommunication or need for repetition implies misunderstanding. For example, K. Renee Horton, a NASA engineer who uses hearing aids, wrote, "I dread having to ask colleagues to repeat themselves because they always rephrase their statement as if I didn't understand its meaning, versus just not hearing the words." [44].

A-4. **Seek out and participate in professional development related to disability and accessibility.** Whether explicit or implicit, bias among instructors and staff is an obstacle to creating an inclusive and accessible learning environment. Instructors may think that, compared to the able-bodied and neurotypical norm, students with disabilities are less capable or intelligent. This assumption can lead instructors to lowering standards for students with disabilities, creating educational inequity. Professional development about disability and accessibility is an important tool for recognizing and overcoming ableist/disablist bias.

A-5.  **Align your actions with your beliefs.** For example, use appropriate vocabulary in the classroom and establish expectations that students will use respectful terminology, too. We discuss some concepts and terminologies in the glossary and the appendices.

## B. Physics education researchers

The physics education research (PER) community plays an important role in establishing standards for undergraduate physics education, compiling and vouching for educational resources (e.g., PhysPort.org), and training physics educators (e.g., through the AAPT New Faculty Workshop). Accordingly, the PER community has an important role to play in improving the accessibility of physics laboratory courses. Below, we outline suggestions for physics education researchers.

B-1.  **Incorporate theories of learning from Disability Studies into physics education research.** Student engagement with apparatus and technology is a main identifying feature of laboratory courses. Such engagement is directly impacted by the course itself, and Disability Studies can provide important insight into interactions between students, apparatus, structures, policies, and practices. For example, the social model of disability suggests that the learning environment can enable some students while disabling others, regardless of whether any students in the course identify as disabled. Traxler and Blue [45] recently wrote a book chapter describing how framings from Disability Studies can be useful in physics.

B-2.  **Prioritize the experiences of students with disabilities within lab courses.** Regardless of specific research agendas, education research on physics labs should ensure that students with disabilities are represented in studies. In addition, there is a need for both qualitative and quantitative research that focuses on the experiences of students with disabilities in lab classes. Examples include:

○ Factors impacting inclusion, exclusion, identity, and sense of belonging in experimental physics and physics more generally;

○ Cultural, affective, and metacognitive strengths of students with disabilities within the laboratory experience;

○ Nontraditional ways of thinking about, experiencing, measuring, or representing physical phenomena.

B-3. **Prioritize research and development of accessible lab courses.** Research and development of lab courses can take many forms. Examples include:

○ Design-based implementation research that incorporates principles of Universal Design for Learning into learning goals, teaching strategies, and apparatus design;

○ Case studies that identify, illustrate, critique, and compare successful instances of accommodating the needs of individual learners in lab courses; and

○ Development of accessible research-based instructional practices for faculty and students in lab courses.

B-4. **Move beyond the traditional researcher-participant dichotomy.** Strategies include using enhanced validation of, or feedback on, research findings from participants via, e.g., synthesized member checking [46]) or implementing research-practice partnerships [47] with practitioners of, e.g., UDL.

B-5. **Consider accessibility when propagating research findings.** Education researchers should consider communicating their research findings in ways that minimize sensory, financial, and geographic barriers. For example, researchers could make their findings more broadly accessible by publishing in open access journals, using arXiv.org to post screen reader-compatible PDF versions of their articles, including alternative-text (alt-text) descriptions of figures, creating video/audio recordings of presentations and posting captioned recordings online, and so on.

C. Physics departments

The ability to correctly anticipate barriers, workload, and time commitments allows students to strategize for success. This is healthier and more efficient than signing up for a one-size-fits-all course plan, only for it to fail and need to be remediated (which costs time, money, and other resources). In addition, for courses that have enrollment caps, all students are impacted when some need to retake a course due to poor planning. The ability to plan ahead creates less stress and more opportunities for everyone.

If students and their support teams know what is expected for each course ahead of time, they can gauge what they/the student can handle, and they can partner with faculty and staff to plan accommodations in advance. The availability of clear and accurate descriptions of lab courses and activities facilitates self-advocacy among students.

C-1. **Regularly discuss which components of lab courses are fundamental.** For each lab course, engage in regular discussions about what would count as a fundamental alteration to the course. Identify which apparatus-based activities are fundamental and how they can be made more accessible. For activities that are not fundamental, proactively discuss the possibility of using simulations, virtual reality, or remote control, sonification, or other strategies to accommodate students for whom the corresponding apparatus is inaccessible. Make course curriculum flexible so that learning differences can be accommodated.

C-2. **Describe each laboratory course comprehensively in course catalogs.** Ensure that prospective students are aware of the learning objectives, workload expectations, technical requirements, types of equipment and software, types of media for instruction and communication, physical and sensory considerations, and types of engagement. Further describe prerequisite knowledge, skills, and coursework comprehensively.

C-3. **Eliminate ableist/disablist language when teaching lab courses and when talking about them.** Ensure that instructors and advisors are able to recognize and avoid ableist/disablist language in the first place. For example, lab instructions could avoid making incorrect assumptions about students' abilities by prompting students to measure or observe rather than to watch or listen. When and if an instructor or advisor becomes aware of their use of ableist/disablist language, they should acknowledge, apologize for, and correct their language in a timely manner.

C-4. **Advise students appropriately.** Ensure that academic advisors have sufficient training to support students to make informed decisions about when to enroll in lab courses based on workload and other cognitive, physical, or sensory expectations. Four-year degree plans are not ideal for all students; some students will need advising that embraces five- or six-year plans. Advisors should encourage students to challenge themselves while respecting boundaries and constraints.

C-5. **Pay for professional development of students, staff, and faculty.** Chairs of Departments and Deans of Colleges should set aside money to pay for opportunities to professionally develop competence with accessibility in physics labs on the part of research faculty, instructional faculty, adjunct faculty, teaching/learning assistants, lab course technicians, members of local Disability Resource Centers, and students with disabilities.

D. Online physics content creators

Physics educators benefit from several online hubs that centralize information about physics teaching practices and research about physics education. Examples include AdvLabs.AAPT.org [48], Compadre.org [49], LivingPhysicsPortal.org [50], PhysPort.org [51], and PICUP.org [52], the first of which is specific to labs and all of which are relevant to labs. These and other websites—including a possible new portal focused exclusively on disability and accessibility—can invest in accessible physics labs in the following ways.

D-1. **Establish and enforce accessibility guidelines for websites and content.** Online physics education portals and the content they host should be accessible to all users. Website administrators and content creators alike should be aware of best practices for accessibility. Accessibility guidelines should be clearly advertised, and enforcement should focus on supporting compliance rather than punishing noncompliance. The World Wide Web Consortium (W3C) created the internationally accepted Web Content

Accessibility Guidelines (WCAG) that establish criteria to make web content perceivable, operable, understandable, and robust [53]. Recently, researchers found all physics departmental websites in the sample ($N = 139$ webpages) had numerous accessibility issues and were in violation of federal law related to website accessibility [54].

D-2. **Create and host blogs, forums, or listservs about accessibility in physics labs.** Online physics education resources could include venues for people to ask questions about accessibility, share ideas, or recommend equipment through blogs, forums, or listservs. The Disabilities, Opportunities, Internetworking, and Technology (DO-IT) listservs are a good example [55].

D-3. **Host or link to resources about accessibility in physics labs.** Resources related to disability and accessibility should be easy to find and up-to-date. Resources can include contact information for appropriate organizations or consultants, links to reports on or guidelines for accessibility in labs, recommendations for implementing Universal Design for Learning in lab courses, examples of specific accommodations, guidelines for effective collaboration with on-campus Disability and Accessibility Offices, etc. Examples that could be emulated or expanded include the websites for the International Association of Geoscience Diversity (IAGD) [56], the American Astronomical Association (AAS) Working Group on Accessibility and Disability (WGAD) [57], and the American Association for the Advancement of Science (AAAS) Entry Point! project [58].

E. Physics conference planners

Conferences play an important role in the professionalization of physics lab instruction. For example, AAPT runs sessions and workshops focused on propagating knowledge or providing training about apparatus and pedagogy in laboratory courses at National Meetings and through the New Physics and Astronomy Faculty Workshop. Additionally, the Advanced Laboratory Physics Association provides similar services through Laboratory Immersions, regional conferences, and the Beyond First-Year Lab Conference. We recognize that some member societies have established working groups or ad hoc committees focused on conference accessibility [31]. For example, in 2018, the AAPT Physics Education Research Leadership Organizing Council (PERLOC) formed a Physics Education Research Conference (PERC) Working Group on Accessibility. They hosted an inaugural Physics Education and Accessible Meeting Planning (PEAMPS) workshop on November 19, 2020. The working group is currently expanding their focus to include AAPT National Meetings in addition to PERC. In and beyond AAPT, conference organizers can invest in accessible physics labs in the following ways.

E-1. **Establish and enforce accessibility guidelines at conferences.** Sessions, panels, workshops, and plenaries should be accessible to all conference attendees. Accessibility oversight committees or consultants can help evaluate and continually improve alignment of conference activities with principles of Universal Design and the requirements of the ADA. Like for websites, accessibility guidelines for conferences should be clearly advertised, and enforcement should focus on supporting compliance rather than punishing noncompliance.

E-2. **Increase awareness of conference-related accessibility issues among organizers.** Increased awareness about accessibility issues can be accomplished by training event staff and volunteers, creating or sustaining a disability and accessibility awareness committee, or ensuring that all committees, forums, working groups, and task forces include at least one person with expertise on accessibility issues.

E-3. **Clearly and publicly articulate conference-related accessibility resources and constraints.** Ensure that prospective attendees are aware of which accommodations will be provided by default, which can be provided upon request, and which cannot be provided due to resource constraints. Considerations include accessible transportation and lodging options, inclusive meeting schedules and room layouts, and availability of interpreters, live captioning options, microphones, screen reader compatibility, maps, and signage. Conference organizers should coordinate with meeting venues to ensure that information is accurate and easily available during the event. All of this information should be made available prior to any registration deadlines and throughout the duration of the conference.

E-4. **Provide grants to defray travel costs for people with disabilities.** Compared to their nondisabled counterparts, individuals who are disabled are disproportionately unemployed or underemployed and often require additional expenses to attend conferences and meetings. Travel grants and other programs can help improve the accessibility of conferences by defraying registration, lodging, transportation, and other incidental costs for attendees with disabilities and, if applicable, their companions.

E-5.   **Solicit post-conference feedback from attendees regarding accessibility.** Any surveys or other post-conference solicitations should be accessible in terms of format and length, and should include information about accessibility at the conference.

## F. Member societies

There is a need for professional development about disability, accessibility, and accommodations for students, staff, and instructors involved with teaching and learning in labs. Many member societies support professional development by organizing conferences or creating online content. For such societies, the recommendations for online physics content creators (Section 3.D) and conference planners (Section 3.E) apply. Additional ideas for member societies are listed below.

F-1.   **Create and advertise professional development opportunities for students, staff, and faculty.** Accessibility-oriented professional development opportunities could strive to achieve the following goals:

- Learn about accessible pedagogies, technologies, and other considerations that are appropriate for different kinds of physics lab courses;

- Support informed self-advocacy among students, staff, and faculty with disabilities;

- Establish a culture of informed allyship among students, staff, and faculty who identify as able-bodied and neurotypical;

- Learn the history of disability justice in the United States and beyond;

- Support all people in recognizing ableism/disablism and to initiate and sustain collective action toward disability justice;

○ Cultivate networks of people with relevant expertise and/or lived experiences with disability and corresponding issues of access and inclusion in physics labs. Networking events could possibly include dialogue with representatives from consulting firms like Independence Science [59]; conferences like the California State University Northridge (CSUN) Assistive Technology Conference [60], or the Inclusion in Science, Learning a New Direction, Conference on Disability (ISLAND) [61].

F-2.  **Increase awareness of ableism/disablism and implications for physics lab courses.** Presentations at national conferences, regional meetings, or departmental colloquiums can highlight issues of ableism/disablism and accessibility in labs. Member societies can further write op-eds, establish social media campaigns, or create accessible web content about these same issues.

F-3.  **Identify and advertise funding opportunities for improving accessibility.** Not all efforts to improve accessibility in physics labs are expensive, but some are. Educators, education researchers, and physics departments may require external funding to implement required changes. To this end, centralizing information about funding opportunities and deadlines is crucial for the advancement of accessibility work in physics labs and physics education more generally.

F-4.  **Purchase assistive devices or accessible apparatus in bulk for reduced per-unit price and sell them at cost to physics departments and individual educators.** Several devices and apparatus have been designed with accessibility in mind; examples include talking LabQuest devices and Snap Circuits®. One way to publicize such devices and apparatus, improve their affordability, and normalize their use would be for member societies to purchase them in bulk, advertise them online, and sell individual units at cost. The Advanced Laboratory Physics Association's Single Photon Detector Initiative [62] is an excellent model of a partnership between a member society and a company with the explicit goal of supporting physics instructors to incorporate new apparatus in physics labs.

**4. Next steps: Make physics labs more accessible!**

After reading this document and some or all of the resources we cite, we hope that readers will reflect, share, collaborate, and take action toward making physics labs more accessible to students, staff, and faculty with disabilities. We anticipate that this document might elicit feelings of discomfort, defensiveness, or guilt alongside motivation, responsibility, or commitment to change. We encourage readers to process any defensiveness they might be feeling and to focus their energy on change.

We hope that readers will share and discuss this document with students, staff, and faculty in their departments, other coworkers, and colleagues in their professional networks. The more broadly the ideas here are shared, the more useful they will be. In addition, by sharing and discussing strategies for making physics labs more accessible, readers will be normalizing conversations about disability, accessibility, and ableism/disablism in physics and other sciences.

Although the ideas outlined in this document are tailored to physics lab instructors and staff, physics education researchers, physics departments, online physics content creators, physics conference planners, and physics member societies, all readers regardless of professional identity may pursue these or other strategies to increase investment in accessible physics labs. Some ideas not explicitly outlined in this paper include establishing or joining independent study courses, departmental committees, professional task forces, or other collaborative efforts. We emphasize that such collaborations must ensure that people with disabilities are given the opportunity and resources needed to participate fully, and that they are offered compensation for their time, effort, and expertise. If sufficient resources exist, readers might consider hiring a consultant with expertise or lived experience with disability [43]. If, in order to achieve goals or implement action items, the collaboration requires training, materials, or other resources, they should ask for them! Universities, professional societies, and other institutions have the responsibility and resources to make physics labs inclusive for students, staff, and faculty with disabilities.

We recognize that our focus on the experiences of students with disabilities in physics lab courses at the undergraduate level results in an incomplete set of recommendations. Additional work is needed to understand and improve the experiences of other stakeholders: staff and faculty with disabilities who support and teach undergraduate physics labs; and people with disabilities who are affiliated with physics labs in K-12 settings, graduate programs, and non-academic research facilities.

**APPENDIX A: Testimonials from current and former students with disabilities**

In the following subsections of this Appendix, we include testimonials from Daniel Gillen, a blind physics bachelor's degree recipient who took on a leadership role in physics lab accessibility at Haverford College, and from three current or former physical science students who are deaf or hard of hearing: Dan Oleynik, Sheila Xu, and Tyler Pugeda.

Testimonial: Daniel Gillen, B.S., Haverford College Class of 2017

The Waves, Electronics, & Optics lab course was one of my favorite experiences during my entire time as a Haverford College physics major. Myself and the College's Office of Access and Disability Services (ADS) had been planning this course for more than a year prior to my enrollment. Each week during the fall term of my fourth year, I would read through the relevant section of the lab manual which I had had meticulously transcribed into Braille with tactile diagrams, studying what was to come in the next class. My in-class assistant and I, along with the lab instructor, would discuss any necessary modifications to original lab manual instructions, at times substituting a more accessible experiment for one that had no viable adaptations.

Some of my fondest memories included constructing various direct current (dc) and alternating current (ac) circuits. For these activities, my other classmates used traditional breadboards with color-coded resistors and signal frequencies in the kilohertz range, examining them visually on an oscilloscope screen. By contrast, my setup included a Snap Circuits® kit with the components labeled in Braille, along with an adapted version of Logger Pro for collecting and sonifying the voltage data. Due to the nature of the sonification, ac signals were limited to frequencies of around 1 Hz, and capacitors were at most 1 mF. It was only for the one experiment on resonance in a resistor, inductor, capacitor (RLC) circuit where we resorted to a more traditional setup. In that case, we lacked a 1 H inductor, which we would have needed in order to achieve low frequencies required for our sonification approach.

Through all of these activities, we as a team always struck a balance between accessibility and what the lab instructor deemed as fundamental within the physics curriculum. If an experiment required visual observation with no viable accessible substitute (which was the case with much of the optics material), we opted for data-focused experiments from the quantum lab course. Whenever I could access and analyze the data from an experiment without visual assistance, I always felt that I was in control of my coursework. As a student who happened to have a disability, I did not let my blindness prevent me from completing my degree or take away my interest in the STEM fields.

Testimonial: Dan Oleynik, Graduate Student, University of Central Florida

I am currently a graduate student at the University of Central Florida, studying physics education research. The impairments that I identify with are being hard of hearing and having to take both anxiety and depression medication for generalized anxiety disorder. The experiences that I have are either treated as "normal" by non-disabled people, and the specifics of both my experience and impairments that I have may not be known.

Instructors regularly misunderstand what it means for me to be hard of hearing. For example, even though I can hear you in one moment, I may not always be able to hear you. This is especially true for popular ways of lecturing, in which an instructor is facing the board or reading from a slide. It's extremely hard for me to hear in this context because the instructor is not speaking in my direction, but speaking somewhere else. While I've gotten good at what I call "translating", it's definitely not perfect.

Being hard of hearing intersects with my anxiety, especially if instructors are not facing towards me, or making their lips visible. Instructors may have to initiate these speaking methods on their own because unless I know the instructor personally, I often feel uncomfortable bringing up my accommodation issue. My discomfort arises because I don't want to distract from other students by putting myself in the spotlight, or I just don't know how to bring it up. For example, in one of the virtual courses I am currently taking, the instructor's audio goes in and out, making it so I can hear them one moment but not the next. I don't know the instructor; I don't know how they'd react, and we're all dealing with COVID anyways, so I don't want to disturb them. So I stopped going to that class. I've tried going to class. I tried engaging. But the mic issue is so bad that it's painful to deal with. In virtual learning environments, having a good mic is good, but having subtitles or closed captioning can also be really helpful. Even if the captions are not great, their presence communicates to me that the instructor is doing at least the minimum to include me. Because I don't want to be the nuisance. And then in terms of my anxiety, I know it can be hard to hear for me, but I'm asking you to not treat me differently, because the last thing I want you to do is treat me differently.

I do understand that instructors are not in control of all aspects of a class, and some accommodations may be possible only if they are accompanied by official documentation. But sometimes, even the bare minimum helps: show understanding, don't call attention to, or make fun of, students who miss assignments, etc. For me, the more of a spotlight that's put on the issue, the more anxiety I feel, which increases the likelihood that I end up putting it off because it becomes a greater mental burden to deal with. Even when accommodations are in place, there are days where I physically can just not focus. There are days where I physically can just not focus on anything. Homework,  research or otherwise. And when those days happen, I just don't turn in homework or assignments, and I call it a 'mental health week.'

Testimonial: Sheila Xu, MIT (B.S., 2014)

For students who use American Sign Language (ASL), hiring a quality sign language interpreter willing to work with them in laboratory courses makes a difference. At MIT, I took a physics-based weather and climate laboratory class that required hands-on experiments. My ASL interpreters had never interpreted for a laboratory course before, much less a laboratory course using physics concepts. At that time, I did not know a lot of signs for physics and weather concepts, and neither did my interpreters. My interpreters and I worked together to come up with signs for both of us that made sense within the context of the experiments. Also, my instructors were very caring and continually checked on me to make sure I understood the concepts demonstrated in the labs. Because of my instructors and interpreters, I had a positive experience in the lab.

Testimonial: Tyler Pugeda, Graduate Research Assistant, M.D./Ph.D. Applicant

An instructor's attitude can make a big difference in my lab experience. Because I was the only deaf person at Johns Hopkins University, I never expected my instructors to be accommodating. On the first day of physics lab, my instructor Dr. Reid Mumford immediately approached me and my American Sign Language (ASL) interpreters in a way that made him seem like a caring instructor. He wanted to understand my access preferences, and he showed a willingness to accommodate me. Surprised at his attitude, I felt comfortable to be in physics lab without feeling the burden of advocating for my needs.

Many experiments later, Dr. Mumford came to my table and informed me that an upcoming experiment would require me to evaluate the quality of audio output from a resistor-capacitor (RC) circuit that was attached to a speaker. I was familiar with the experiment, but I appreciated that Dr. Mumford anticipated my potential challenge of not accessing sounds. He assured me that he would be with me during the experiment and that he would make sure that I had all the necessary information to complete my lab report. Despite supervising eight lab sections at once, Dr. Mumford took time to invest specifically in my education.

On the day of the experiment, my lab partner and I used Snap Circuits® to construct our RC circuit. To analyze the circuit's behavior, we connected an oscilloscope to the speaker to visualize the output. With the aid of my ASL interpreters, my partner and I heard and recorded different sounds (e.g., treble or bass) and their intensities. For example, treble signals became more clear when we increased the capacitance of the capacitor. Dr. Mumford answered my questions and clarified the experimental process. Although the experiment was designed to accommodate hearing students, I appreciated that Dr. Mumford made sure that I was not excluded from experiments or the lab space. Because of his welcoming attitude, I had a positive experience in physics lab.

**APPENDIX B: Autistic learners**

Autism, which is officially referred to as ASD[1] by the DSM-V, represents a diverse set of neurodevelopmental conditions that affect communication, behavior, and the experiences of one's surroundings from an early age, and it affects people of all races and genders.  While previously categorized separately, Autism and Asperger's syndrome are now considered part of a broad non-linear 'spectrum' due to their variability, and recent estimates indicate about 2.2% of the U.S. adult population ages 18 and older lives with Autism [63].

Symptoms and struggles present differently in each person, with people of all ages experiencing mild to severe challenges that can vary over their entire lives. Autistic individuals may experience difficulties with social expectations and communication, as well as compulsive or repetitive behaviors or interests that sometimes manifest as fixations. Autism often manifests with other chronic conditions, typically referred to as comorbidities or co-occurring conditions, such as sensory processing issues (SPD), dyslexia, motor and speech difficulties, attention-deficit/hyperactivity disorder (ADHD), and obsessive-compulsive disorder (OCD).

Appropriate language and discourse around Autistic individuals has changed significantly over the last several decades.  The Autistic community is another Disabled group for whom a majority currently prefers identity-first language over person-first, but similar to asking for one's gender pronouns, it is appropriate to use the language that an individual prefers.

Problematic labels such as "high-functioning" and "low-functioning" have historically been associated with certain symptoms of Autism, and their use can be harmful [64].  These labels often coincide with how 'productive' the person is to a capitalistic and materialistic society.  Labeling can reinforce stereotypes, such as the incorrect assumption that nonverbal people necessarily have low cognitive ability. It can also present social barriers to an individual seeking assistance, such as assuming they need to 'try harder' to be 'normal' and are 'just being lazy'.  Similarly, while some Autistic individuals do have academic areas or traits in which they excel, enforcing expectations on Autistic individuals to be geniuses or 'savants' can be harmful and should be avoided, just as one would avoid perpetuating other fixed-mindset attitudes in student learning environments.

---

[1] ASD stands for Autism Spectrum Disorder.  However, many members of the Autistic community oppose use of the word "Disorder" with regards to the Autism Spectrum.  For more on how to talk about Autism without using the word 'Disorder', see the following page from the Autistic Self-Advocacy Network ASAN, an organization run by Autistic people: https://autisticadvocacy.org/about-asan/about-autism/

The way educators interact with Autistic learners has lasting impacts on those learners. College introduces many new stressors, in particular significant increases in unstructured learning and social interactions. Attending college may also be the first time that Autistic students are navigating accommodations on their own without the guidance of a parent or guardian. It is important for faculty and staff to foster trust and safety in educational settings by communicating clearly and directly, and to be cognizant of the number of stereotypes and misconceptions around Autistic people— many of which are outdated and potentially harmful.

Many Autistic individuals have developed physically and mentally taxing coping strategies to hide their Autistic characteristics and behaviors and appear neurotypical. This behavior is called 'camouflaging', and it includes 'masking', or 'assimilation.' With roots in survival and social acceptance, camouflaging often is seen beginning to develop in early childhood, with significant variation depending on assigned gender [65]. While masking can be unconscious, it can also be a conscious choice or coping mechanism. Many times, masking can be forcibly imposed by parents or trusted authority figures, who may call out or otherwise punish conspicuous Autistic behaviors, such as stimming[2], speaking thoughts out loud, and avoiding eye contact.

Autism does not magically disappear when a person turns 18. Nevertheless, for many adults, obtaining a formal diagnosis—which in turn provides avenues to mandated formal accommodations and other assistance—can be impractical or prohibitive for a variety of reasons, including cost, time, and emotional energy. Moreover, successfully masking Autistic traits does not 'cure' or mitigate an Autistic person's symptoms. It is also important to note that gender differences in extent of masking is considered to be one reason that the rate of diagnosis of Autism skews male; a significant number of those with female socialization and upbringing remain formally undiagnosed, or are not diagnosed until adulthood. Persistent systemic racism and socioeconomic class may additionally account for significant disparities in Autism diagnosis for BIPOC[3] individuals, who are more likely to be misdiagnosed with violent or disruptive behavioral conditions [66], [67]. Increasing access for neurodiverse community members thus has intersectional impacts far beyond formally documented accommodations needs.

---

[2] Stimming, or stereotypy, is behavior that accomplishes controlled stimulation of senses in order to manage external sensory overload. Common stims can include rocking back and forth or fidgeting, spinning toys, or repeating phrases. Stimming is a valuable behavior to many Autistic people, as it can be a method for emotional regulation, and can be a defensive strategy against more severe consequences of sensory overload such as meltdowns.

[3] BIPOC stands for Black, Indigenous, and People of Color  https://www.thebipocproject.org

The Autistic authors of this appendix recommend that those interested in further reading preferentially consult resources provided by the Autistic community that center on Autistic lived experiences, and to be cautious of resources and websites championed by non-autistic individuals and organizations. Websites with good beginners' resources include:

- The Autistic Self-Advocacy Network (ASAN)
- The Autistic Women and Nonbinary Network (AWN)

**APPENDIX C: Blind, visually impaired, and low vision learners**

Legal blindness in the United States includes anyone whose visual acuity is 20/200 or less, or whose visual field is 20 degrees or narrower. For this reason, students who are legally blind span a wide range of visual acuities, and may even have 20/20 vision, but with a narrow visual field. For the sake of convenience, we will use BVI to refer to blind or visually impaired students, a group comprising those with various levels of vision loss. Terms in common use include "visually impaired," "low vision," "partially sighted" (mostly outside the U.S.), "legally blind", and "totally blind," with or without light perception.

The amount of functional vision a person has is often not a reliable predictor of which terminology they prefer when describing themselves. It is best to ask them about such preferences, as opposed to assuming how much vision (useful or not) a person has, or drawing comparisons between students who are blind based on such assumptions. A student who seems to have usable vision "on paper" may experience photosensitivity, fatigue, or debilitating migraines when required to use their sight for prolonged periods, and therefore may opt for non-visual means of accessing information. Another student with seemingly similar causes of blindness may have vastly different preferences. When discussing accessibility solutions and accommodations with students who are blind, it is important to focus on the desired outcome, and not on maximizing available vision or on replicating the same process being performed by other classmates. The same is true when considering matters of lab safety. In all cases, it is important to continually seek the blind student's input, prior to, throughout, and after the lab course.

According to the 2018 National Health Interview Survey (NHIS), "an estimated 32.2 million adult Americans (or about 13% of all adult Americans) reported they either 'have trouble' seeing, even when wearing glasses or contact lenses, or that they are blind or unable to see at all" (Statistical Snapshots from the American Foundation for the Blind). Among science and engineering doctorate recipients, the National Science Foundation (NSF) found that, in 2017 alone, more than 1,200 people (or about 3.5% of all science and engineering doctorate recipients) reported having visual disabilities (NSF National Center for Science and Engineering Statistics). Thus, BVI students encompass a significant proportion of the student population.

In addition to focusing on course outcomes when discussing accommodations with a BVI student, it is important to consider the role of effective communication in labs. For BVI students, communication barriers relate mostly to inherently visual methods of teaching or presenting physics material in class. Despite having no or low vision, BVI students have great potential to excel in physics and other STEM subjects when provided with appropriate support and accommodations.

**APPENDIX D: Deaf and hard of hearing learners**

When we refer to "deaf and hard-of-hearing," we also include Deaf, DeafBlind, DeafDisabled, Hard of Hearing, Late-Deafened Students (DDBDDHHLD), and other groups of varying hearing levels to respect everyone's unique identity and background. Please note the words "hearing-impaired," "deafmute," "deaf and dumb," and other derogatory words are not appropriate to use for the purposes of political correction.

Deaf and hard-of-hearing (DHH) students are unique. Unlike hearing students with disabilities, communication barriers are inherent for DHH students. There is a varying range of hearing loss and a diversity of different communication methods unique to each DHH student. For example, some DHH students require an American Sign Language interpreter, others need captioned videos, and some need assistive hearing devices. There are also students who don't use sign language and only rely on speaking and listening. In the US alone, there are at least 2 million deaf Americans, so DHH students are a significant group of students with disabilities in educational institutions across the country.

We must invest in supporting DHH students to ensure that they receive the same quality of physics education as their hearing counterparts. A young, budding physicist in today's classroom could be a DHH student with a potential interest and promise for a career in physics if we take the right actions today to invest in their accessible physics education. We want these students to become well-informed, educated citizens with a solid understanding (and perhaps, a new love) of physics - for the benefit of our multicultural society and country.

**APPENDIX E: Glossary of terms**

This glossary includes definitions of select disability terminology. The list of terms is neither authoritative nor exhaustive. For example, the Disabled People's Association, Singapore (DPA) has published a much longer glossary of disability terminology [68] with distinct definitions of some of the terms that appear here. In addition, UW DO-IT maintains an online glossary of disability-related terms [69], and Autistic Hoya has published a glossary of ableist phrases [70].

Groups and individuals determine the best terminology to use, and there may not be consensus among all individuals or groups. Finally, terms change over time; some terms become outdated or take on different meaning as time passes, and new terminology gets created.

**Ableism:** Bias or discrimination in favor of able-bodiedness or against persons with one or more disabilities, or a belief that those without disabilities are an expected societal norm. Ableism can occur in forms ranging from inaccessible physical or virtual spaces to attitudes or customs that favor people without disabilities [71], [72]. *See also: disablism.*

**Accessibility:** Usability by or suitability for people with disabilities. In this sense "[t]he concept of accessible design and practice of accessible development ensures both "direct access" (i.e. unassisted) and "indirect access" meaning compatibility with a person's assistive technology (for example, computer screen readers)." [73].

**Accommodation:** A physical, technological, or methodological modification to an activity that aims to allow people with disabilities to enjoy equal opportunities in engaging in said activity. In the context of physics lab courses, students with disabilities may, for example, require adapted computer hardware/software or lab equipment, or one or more assistants (note-takers, sign language interpreters, etc) to make the course accessible to them [74], [75].

**Allyship:** Work done by those who are not under-represented in one community (e.g., neurotypical/non-disabled with regards to the disability community) in order to support the marginalized people and groups within that community.

**American Sign Language (ASL):** The predominant sign language used by Deaf communities in the U.S. and English-speaking regions of Canada [76].

**Americans with Disabilities Act (ADA) of 1990:**  A comprehensive piece of U.S. civil rights legislation that mandates equal protection, access, and opportunities to individuals with disabilities [77], [78].

**Autism Spectrum (ASD)**: A diverse range of neurodevelopmental conditions including Autism and Asperger syndrome, among others, that affect communication and behavior from an early age; it presents differently between individuals. Common characteristics include difficulties or impairments in social interactions and communication, as well as repetitive, restrictive, or compulsive behaviors and interests [79]. *See also: Appendix B*.

**Bias (Explicit):** Attitudes and beliefs that we consciously hold about a person or about a group of people. Explicit biases can be both positive or negative.

**Bias (Implicit):** Attitudes and beliefs that we unconsciously hold about a person or about a group of people. Due to the systems that we participate and are socialized in, each person has conditioned implicit biases that we continually work to acknowledge and address.

**Bias-free language:** An inclusive writing and communication style that uses composition and word choice free of prejudicial language.  Bias-free language avoids using implicit or normative assumptions of the reader's age, race, ethnicity, gender, sexuality, disability, socioeconomic status, religion, or intersectionality, as an example [70], [80], [81].

**Braille:** A tactile writing system used by blind and visually impaired people worldwide. Characters are formed through combinations of six raised dots arranged in 2×3 rectangular arrays. Braille is not a language, but a way of representing a print-based written language or other writing system. Hundreds of regional and national Braille codes exist around the world, some of which can represent mathematical and scientific notation (i.e., Nemeth Code in most English-speaking countries).

**Comorbidity**: The simultaneous occurrence of two or more chronic conditions or diseases in an individual.  Many disabilities have comorbid conditions associated with them.

**Disablism:** Bias or disrimination against disability and people with disabilities. "Disablism relates to the oppressive practices of contemporary society that threaten to exclude, eradicate and neutralise those individuals, bodies, minds and community practices that fail to fit the capitalist imperative." [82] *See also: ableism.*

**Disability (medical model):** A physical or mental impairment that substantially limits an individual to carry out one or more normal life functions. In this model, disability is situated within the individual. This (standard) medical model "supposes that this disability may reduce the individual's quality of life and the aim is, with medical intervention, this disability will be diminished or corrected." [83]**.**

**Disability (social model):** Systemic barriers, derogatory attitudes, and intentional or unknowing social exclusion make it difficult or impossible for people with disabilities to achieve life goals. This view diverges from the dominant medical model and situates disability within the environment that is planned for able-bodied people. Yet "[w]hile physical, sensory, intellectual, or psychological variations may cause individual functional limitation or impairments, these do not necessarily have to lead to disability unless society fails to take account of and include people regardless of their individual differences." [84]**.**

**Disability Studies:** An interdisciplinary academic approach to studying the medical, social, and sociological aspects of disabilities through the humanities, sciences, and social sciences. "Programs in Disability Studies should encourage a curriculum that allows students, activists, teachers, artists, practitioners, and researchers to engage the subject matter from various disciplinary perspectives." [85].

**Equality:** The Center for the Study of Social Policy defines equality as "The effort to treat everyone the same or to ensure that everyone has access to the same opportunities. However, only working to achieve equality ignores historical and structural factors that benefit some social groups and disadvantages other social groups in ways that create differential starting points." [86]

**Equity:** The Center for the Study of Social Policy defines equity as "The effort to provide different levels of support based on an individual's or group's needs in order to achieve fairness in outcomes. Working to achieve equity acknowledges unequal starting places and the need to correct the imbalance." [86]

**Fundamental alteration or change:** When making a change to a good or service in order to accommodate the needs of an individual or a group, a fundamental alteration refers to a change that is so significant that it changes the scope or original goods or services being offered. (e.g., if a change significantly alters the purpose or educational material of a class being taught, that would be a "fundamental alteration.") It highlights a limit to the ADA as businesses are not legally required to change if it would cause 'fundamental alterations.' In educational settings, fundamental alterations of the learning environment are typically referred to as "course modifications." [87] *See also: Accommodation, Inclusion.*

**Identity-first language:** The linguistic practice of mentioning an aspect of an individual's identity (particularly disabilities or other medical conditions) before any nouns related to personhood. Examples include "visually impaired students," "Autistic child," and "Deaf person." Often contrasted with person-first language, identity-first language is strongly preferred within a number of Disabled communities. The preferred use of identity-first over person-first language in bias-free language has been rapidly evolving as of the compilation of this glossary [88], [89]. *See also: Person-first language*

**Inclusion:** A design philosophy for educational, occupational, or other environments in which diversity in a community is respected. In essence, all members of a community are treated without pity or judgment for being part of groups traditionally marginalized by society. A lab course designed with inclusion in mind should thus embrace the high potential of students with disabilities to succeed, in addition to respecting differences in race, gender identity, or other legally protected categories. Inclusive teaching strategies are instructional practices that are implemented to proactively design learning environments to support the diversity of learners who may enroll in the course. Inclusive teaching strategies align with UDL. *See also: Universal Design for Learning.*

**Individuals with Disabilities Education Act (IDEA) of 1975:** A U.S. Federal law that requires equal access to free public K-12 education for eligible children with disabilities. In addition, "[t]he IDEA governs how states and public agencies provide early intervention, special education, and related services" to these children [90]. IDEA applies at the K-12 level. See also: Americans with Disabilities Act (ADA) of 1990 and Rehabilitation Act of 1973.

**Masking**: Also known as camouflaging. The obscuring of features that define a person's identity, especially as it pertains to disabilities. Hiding or camouflaging a disability is either a conscious choice or subconscious reaction in order to better fit normative or neurotypical environments and social expectations. The term masking is most often used within the Autistic community.  Masking is often taxing on the individual, and can be associated with negative long-term harm.

**Neurodiversity**: "[A] concept where neurological differences are to be recognized and respected as any other human variation." This concept is compatible with the social model of disability, and challenges the idea that neurodevelopmental differences are inherently pathological [91], [92].

**Neurotypical**: A term that may be used alongside discussions of neurodiversity in the Autistic community.  It is used by Autistic people to refer to abled individuals who are not on the Autistic spectrum, or to non-Autistic traits, or "those with strictly typical neurology, that is, without a learning disorder or neurodevelopmental disorder." [93].

**Person-first or people-first language:** The linguistic practice of mentioning a person or population before an aspect of their identity (often a disability or other medical condition), for example "students with disabilities," "person with a visual impairment," or "student who is Deaf". It may describe what a person "has" rather than assert what a person "is." "It is intended to avoid marginalization or dehumanization … when discussing people with a chronic illness or disability", but is becoming increasingly disfavored in writing as language around Disability evolves.  Often contrasted with (but may also supplement) identity-first language [94]–[96]. *See: Identity-first language.*

**Rehabilitation Act of 1973:** A U.S. Federal civil rights law that "prohibits discrimination on the basis of disability" in the context of programs and employment as part of or funded by the Federal government. It was the basis for the Individuals with Disabilities in Education Act (IDEA) and the Americans with Disabilities Act (ADA) passed in 1975 and 1990 respectively. "The standards for determining employment discrimination under the Rehabilitation Act are the same as those used in title I of the Americans with Disabilities Act." [97]. Sections 504 (prohibiting discrimination in areas of postsecondary education functions) and 508 (requiring access to electronic and digital information) are particularly relevant to the postsecondary context.

**Screen reader:** A piece of assistive software that allows the contents of a computer, tablet, or smartphone screen to be read aloud with synthetic speech and/or displayed on a refreshable Braille display. Some screen readers are directly implemented in mainstream operating systems (e.g., VoiceOver in Apple's iOS or MacOS), while most others are paid programs from external companies (e.g., Freedom Scientific's JAWS for Windows).

**Universal Design for Learning (UDL):** "UDL is a framework to guide the design of learning environments that are accessible and challenging for all. Ultimately, the goal of UDL is to support learners to become "expert learners" who are, each in their own way, purposeful and motivated, resourceful and knowledgeable, and strategic and goal driven. UDL aims to change the design of the environment rather than to change the learner. When environments are intentionally designed to reduce barriers, all learners can engage in rigorous, meaningful learning." [98]. Interested readers can learn more about UDL in *Universal Design for Learning: Theory and Practice* by Meyer and Rose, Center for Applied Special Technologies [99] or in Universal Design: Places to Start by Jay Dolmage [100]**.**